\documentclass{CUP-JNL-DAP}
\usepackage{graphicx}
\usepackage{multicol,multirow}
\usepackage{amsmath,amssymb,amsfonts}
\usepackage{mathrsfs}
\usepackage{amsthm}
\usepackage{apacite}
\usepackage{rotating}
\usepackage{appendix}
\usepackage[authoryear]{natbib}
\usepackage{ifpdf}
\usepackage[T1]{fontenc}
\usepackage{times}
\usepackage{newtxmath}
\usepackage{textcomp}
\usepackage{xcolor}
\usepackage[bookmarks=false]{hyperref}

\articletype{Research article: preprint, accepted for publication in Data\&Policy Journal}
\jyear{Dec 2023} 
\jdoi{xx.xx/xx}
\begin{document}

\begin{Frontmatter}

\title{On the use of smart hybrid contracts to provide flexibility in algorithmic governance}

\author*[1]{Carlos Molina--Jimenez}\email{carlos.molina@cl.cam.ac.uk}\orcid{0000-0002-3617-8287}

\author[2]{Sandra Milena Felizia}\email{feliziasandra19@gmail.com/SandraMilena.Felizia@UAI.edu.ar}\orcid{0000-0000-0000-0000}
\authormark{Carlos Molina--Jimenez}

\address*[1]{\orgdiv{Department of Computer Science and
 Technology}, \orgname{University of Cambridge}, \orgaddress{\city{Cambridge},\street{15 JJ Thompson Ave.}, \postcode{CB3 0FD},\country{England}}}

\address[2]{\orgdiv{Facultad de Ciencias Econ{\'{o}}micas}, \orgname{Universidad Abierta Interamericana}, 
            \orgaddress{\city{Rosario}, \postcode{2000}, \state{Santa Fe}, \country{Argentina}}}

\received{dd mm yyyy}
\revised{dd mm yyyy}
\accepted{01 12 2023}


\keywords[Abbreviations]{FSM, Finite State Machine; HC, Hybrid contract.}

\abstract{The use of computer technology to automate the 
  enforcement of law is a promising alternative to simplify 
  bureaucratic procedures. However, careless automation might 
  result in an inflexible and dehumanize law enforcement 
  system driven by algorithms that do not account for the 
  particularities of individuals or minorities. In this paper, 
  we argue that hybrid smart contracts deployed to monitor 
  rather than to blindly enforce regulations can be used to 
  add flexibility. Enforcement is a suitable alternative only 
  when prevention is strictly necessary; however, we argue that 
  in many situations a corrective approach based on monitoring 
  is more flexible and suitable. To add more flexibility, the 
  hybrid smart contract can be programmed to stop to request 
  the intervention of a human or of a group of them when human 
  judgement is needed.}

\begin{policy}
 The article assumes that algorithmic governance will be gradually adopted
by governments, which implies that we are heading to a society where the law
is enforced automatically by computer-executable programs called smart contracts
(digital contracts, programmable contracts, etc.). The authors argue that
smart contracts are inflexible, likely to suffer from gaps and, more
importantly, lack human judgment. Therefore, there is a risk of creating 
a de--humanize law enforcement system driven by algorithms that do not account 
for the particularities of individuals or minorities. To address the problem,
they suggest that smart contracts should work in tandem with humans to 
be involved in situations where human sense is needed.
\end{policy}

\end{Frontmatter}

\section{Introduction}
\label{introduction}
 The strong relationship between logic and law has been 
 acknowledged since ancient times and the subject of
 interest for decades~\cite{Darrel1964}. It is widely
 acknowledged that regulations, at least partially, can be
 modelled by logical statement like \emph{if event\_occurs and
 condition\_holds then execute\_action} that can be 
 expressed as computer code that can be executed mechanically.
 For example, \emph{if transaction executed and amount larger than 10000
 then report to government}.
 
 Recent progress in computer technology has generated excitement
 about the possibility of building systems that enable
 law automation. This is a recently emerging concept referred 
to by different terms. In~\cite{PrimaveraAaron2018} it is 
called Lex Cryptographia.
in~\cite{ComLawMIT2021,MichaelNathaniel2005,MichaelHomePag} it is 
called Computational Law, in~\cite{James2017,Eva2020} it is 
called Replacement of Law with Computer Code, in~\cite{Harry2012} 
it is called Computable Contracts. It
can fairly be called Programmable Law because it is a law 
implemented by computer programs or Algorithmic 
Governance~\cite{Urs2022,Marta2021,Kevin2020} to emphasise that 
law enforcement will follow algorithms, that is, strict 
mathematical procedures.
 
The general idea behind all these terminologies is to use computer code 
to automate the enforcement of regulations in different fields of 
our society, ranging from regulations within private companies to 
governments. This would require translating laws (civil code, 
codes covering corporate law, administrative law, tax law, constitutional 
law, etc.) which is currently written in natural language for human 
interpretation, into a code that computers can read, interpret and 
execute automatically. In addition, this code needs to be protected 
against accidental and malicious threats. Some authors use the term 
smart contract to refer to this and similar code that can be used for
the automation of regulations.

We can define a smart contract (also known as digital contract, executable 
contract and automatic contract) as a piece of executable
computer code that a software engineer implements from the translation
of normative statements written in natural language into a computer
language such as Python, Solidity, Go, etc.

 Current governments are infamous for their cumbersome and unnecessarily slow
 bureaucratic procedures; for example, it takes months or even years for  
 courts to dictate a sentence and involve scores of 
 printed documents. Fortunately, algorithmic governance promises to simplify 
 and speed up bureaucratic procedures. More importantly, the adoption of 
 computer technology by the legal system opens opportunities to ameliorate 
 the drawbacks that afflict current political systems, such as minority 
 exclusion~\cite{Peter2018}. However, algorithmic governance raises new 
 challenges\cite{SandraCarlosRafa2022}. In our opinion, one of the most 
 important challenges is the adoption of automatic preventive laws, as 
 we explained in Section~\ref{futurework}.

 In this work, we raise the question about the lack of flexibility 
 that algorithmic governance can potentially introduce. There are concerns 
 that smart contracts are inflexible~\cite{JeremySkla2018} 
 software mechanisms, consequently, their use in the implementation 
 of computational law would compromise the flexibility of 
 the current law which is based on the intervention of humans 
 to provide human judgment. To ameliorate
 the problem, we suggest the use of an  incomplete hybrid smart 
 contracts that can be deployed to monitoring and 
 enforce as necessary, rather than only to mechanically 
 enforce regulations. 
 In addition, we suggest that in border--line situations
 where human judgment is needed, the incomplete hybrid smart contract 
 stops to request human intervention. The response can be produced by
 a single individual or by a group of them after reaching a  consensus.
 
 Human judgment is needed where the decision to be taken would 
 have an irreversible effect on an individual or society. At the 
 top of the list, we would place situations that have been documented to be
 challenging to handle with computer technology. For example, it has 
 been widely documented that algorithm--based image recognition is not 
 reliable; therefore, it is too risky to use medical images in 
 life--threatening surgeries without human examination for final 
 approval. This and other examples of situations where computers 
 fall short and therefore need human help are discussed 
 in~\cite{CharlesChoi2021}.

 The rest of this paper is developed as follows: In
 Section~\ref{contractrop} we introduce concepts
 related to normative statements, including
 contracts, rights, obligations and prohibitions.
 In Section~\ref{contexecution} we explain how
 an automatic contract can be deployed for
 monitoring and enforcement. In Section~\ref{hybridcont}
 we explain the solution that we suggest for providing
 flexibility in algorithmic governance. In Section~\ref{futurework} 
 we discuss preventive law as a future research topic. We suggest 
 that automatic preventive law can be used as a measure to prevent 
 the execution of criminal acts, as opposed to criminal punishment. 
 For example, it can be used to deter monopolistic practices.   
 In Section~\ref{concluding} we close the discussion with 
 some remarks that reflect lawyersr' concerns about the
 invasion of computer technology of a field that has been for 
 centuries lawyers' exclusive domain.

\section{Contracts, rights, obligations, prohibitions and operations}
\label{contractrop}
From a technical perspective, a contract is conceived as a set
of clauses that stipulate \textbf{rights}, 
\textbf{obligations} and \textbf{prohibitions} 
that the signatories are expected to comply with rights, 
obligations and prohibitions are associated
with at least one operation. An \textbf{operation} is a business
action executed by a party that changes the state of the
contract development, for example, pay bill, deliver
item, etc.
 
In simple contracts, each right, obligation and prohibition is associated 
with a single operation. In these situations, one can regard a right as an
operation that a party is entitled to execute. Likewise, an obligation 
can be regarded as an operation that a party is expected to execute. 
Finally, a prohibition can be regarded as an operation that 
a party is not expected to execute.
As an example, we can think of a contract where
Bob has the obligation to pay Alice 100.00 under certain conditions 
(by 31/Dec/2020). To honour this obligation, Bob 
needs to successfully execute through some mechanism 
the corresponding operation ''pay 100.00 to Alice'' by the deadline.
In practice, contracts include several obligations. For example,
a buyer has the obligation to pay and a seller the obligation
to deliver or the obligation to refund. Therefore, to comply with the 
whole contract, the signatory parties need to honour each 
obligation by means of executing the corresponding operation ---
pay, deliver and refund in this example. 
The motivation for using digital contracts is that they
automate the execution of operations in compliance with the rights, 
obligations and prohibitions stipulated in the contract. Automatic 
execution frees the signatory parties from the hassle of performing 
them manually to honour the corresponding obligations.

To understand the enforcement of a whole digital contract, it helps to
regard the execution of a digital contract as the execution of several 
interrelated operations where the execution of one of them exercises a right, 
honours an obligation or violates a prohibition and might enable or disable other
rights, obligations and prohibitions. Some authors regard each right, obligation 
and prohibition as an individual contract. In their model, a 
contract is composed of one or more interrelated subcontracts.

\section{Contract execution}
\label{contexecution}
\begin{figure}
\centering
\includegraphics[width=0.65\columnwidth]{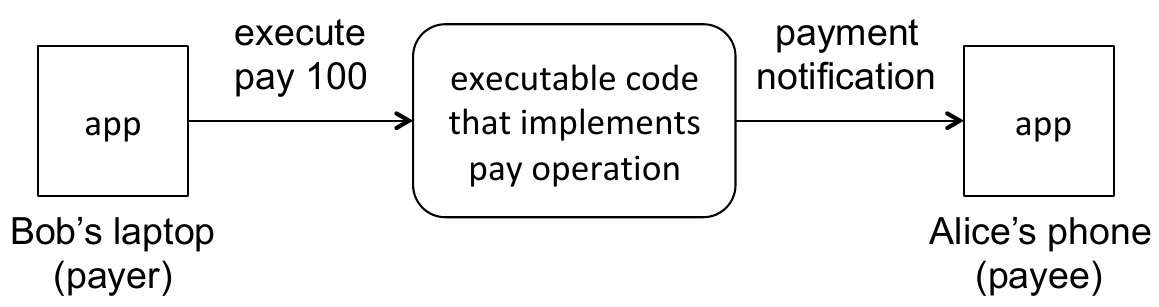}
\caption{Execution of pay operation without the involvement of a smart contract}
\label{fig:exepayoper}
\end{figure}

In automatic contracts, operations are executed through the
execution of the executable code that implements them. 
Fig.~\ref{fig:exepayoper} shows the execution of a payment 
operation. Pay 100 is assumed to be stipulated in a contract agreed 
upon between Bob (the payer) and Alice (the payee). Bob's application 
is installed on his laptop and Alice's is installed on her mobile 
phone. The executable code that implements the pay operation is assumed 
to be implemented in a programming language and deployed in a computer, 
a local one, in a cloud server or on a blockchain. 
To illustrate the architecture that we suggest
in this paper for providing flexibility 
(Fig.~\ref{fig:hybridcontconsensus}) we will assume that
the contract is deployed on a blockchain, for example 
on the Ethereum blockchain~\cite{EthereumHome}.

In principle, smart contracts can be executed in centralized 
systems (for example, Carta \footnote{https://carta.com/}) following 
the traditional client-server model which is far simpler and 
better understood than the other 
alternative is the use of decentralized systems like blockchains. 
We acknowledge that blockchains are far more complex and still 
under test. However, we argue that in some situations, the advantages 
that decentralized systems bring, outweigh complexity. For 
example, in some government applications (say, budget expenditure), 
transparency, traceability and indelible records are essential 
requirements. These properties are naturally provided by blockchains 
and are difficult to implement in centralized systems.

Returning to the example of Fig.\ref{fig:exepayoper}, to pay 
Alice, Bob's application issues the operation ''pay 100'' 
against the executable code. As a response, the executable code 
executes the operation and as a result, Alice's application receives a 
notification of payment, for example, bank evidence of the 
payment. Notice that Fig.~\ref{fig:exepayoper} shows only the execution
of the pay operation. There are no digital mechanisms in place to 
detect Bob's failure to execute the pay operation or to enforce 
him to execute it automatically. Automatic enforcement can be
achieved with the help of a digital contract.
 From the perspective of the level of interference that 
 the digital contract causes in the execution of the 
 contractual operations, digital contracts can be deployed to 
 either monitor or enforce. The two alternatives are shown, 
 respectively, in Fig.~\ref{fig:exepaymonitor}
 and Fig.~\ref{fig:exepayenforce}.
 The advantage of automatic enforcement and monitoring is fundamental in 
 law automation. Unfortunately, existing literature focuses only on
 enforcement and fails to appreciate the advantages of monitoring. See for 
 example Lex Cryptographia~\cite{PrimaveraAaron2018}.

\subsection{Contract monitoring}
\label{contmonitoring}
Contract monitoring is a technique where a smart contract
is deployed to observe the development of the action
passively and to store records of the operations executed by the
signatory parties. Monitoring is passive in the sense that
the smart contract does not interfere with the development of the 
action; it only observes and keeps records for potential 
post--mortem examination, for example, if a dispute is raised.

\begin{figure}
\centering
\includegraphics[width=0.72\columnwidth]{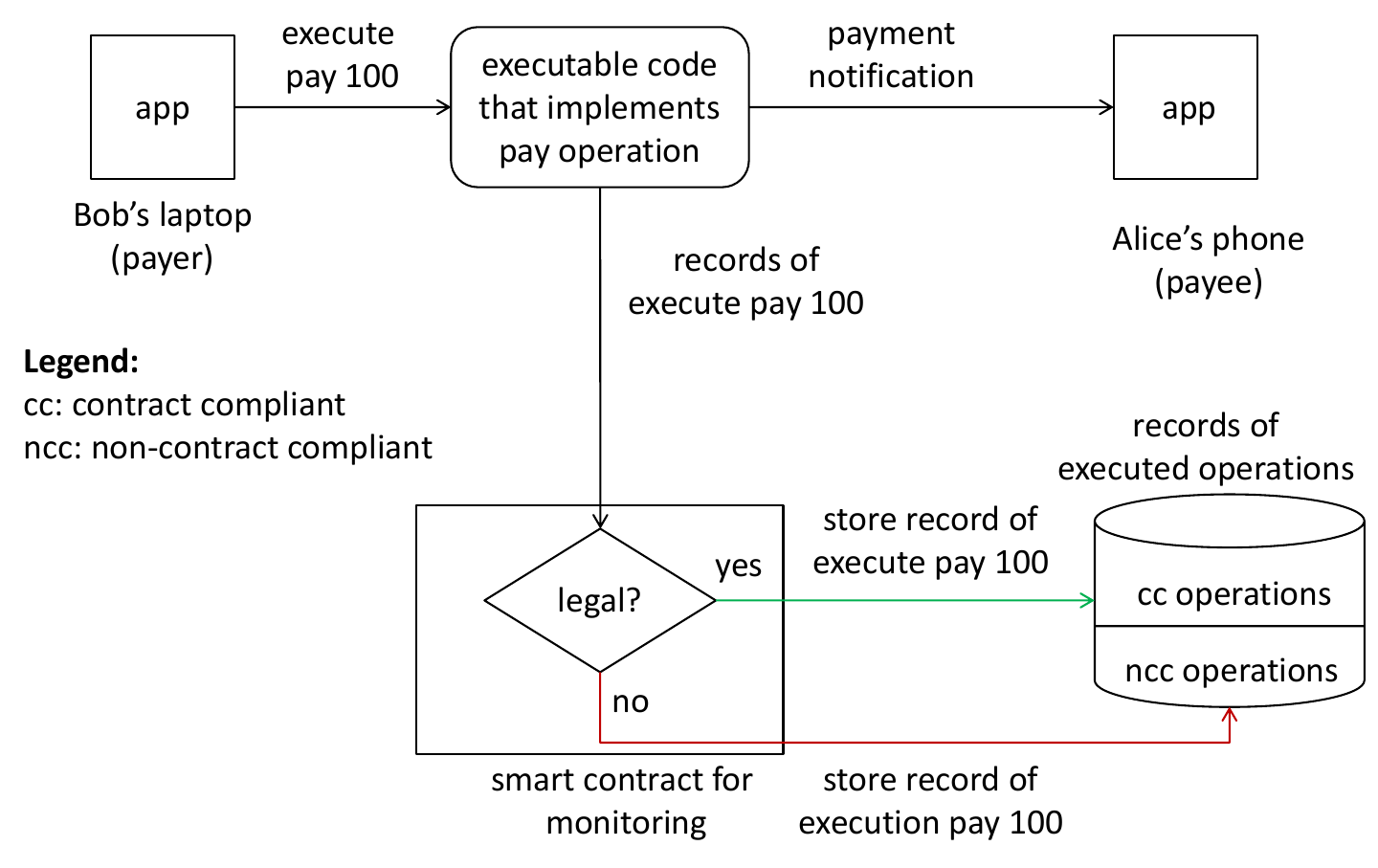}
\caption{Monitoring of the execution of a pay operation}
\label{fig:exepaymonitor}
\end{figure}

Fig.~\ref{fig:exepaymonitor} shows how a smart contract can be
deployed for monitoring the  execution of a payment operation. 
Notice that the smart contract is directly interrelated with 
executable code that implements the payment operation. In fact, 
in existing literature, the two components are frequently discussed as
a single one. We separate them to help understand how smart 
contracts work.

\begin{enumerate}	
\item Bob's application places the operation ''execute pay 100'' 
 against the executable code.

\item The executable code executes the operation and as a 
  result,  Alice's application receives ''payment notification'',
  for example, a bank receipt.

\item The executable code provides the smart contract
  that is responsible for monitoring  with records of the execution 
  of the pay 100 operation placed by Bob.
  
\item The monitoring contract analyses the records,
  determines if ''pay 100'' operation is contract
  compliant (legal) or non--contract compliant and sends
  its verdict (red and green lines, respectively) to the
  database. The records accumulated in the
  database can be used for conducting post--mortem 
  (off--line) examination of the contract development. 
\end{enumerate}

\subsection{Contract enforcement}
\label{contenforce}
\begin{figure}
\centering
\includegraphics[width=0.78\columnwidth]{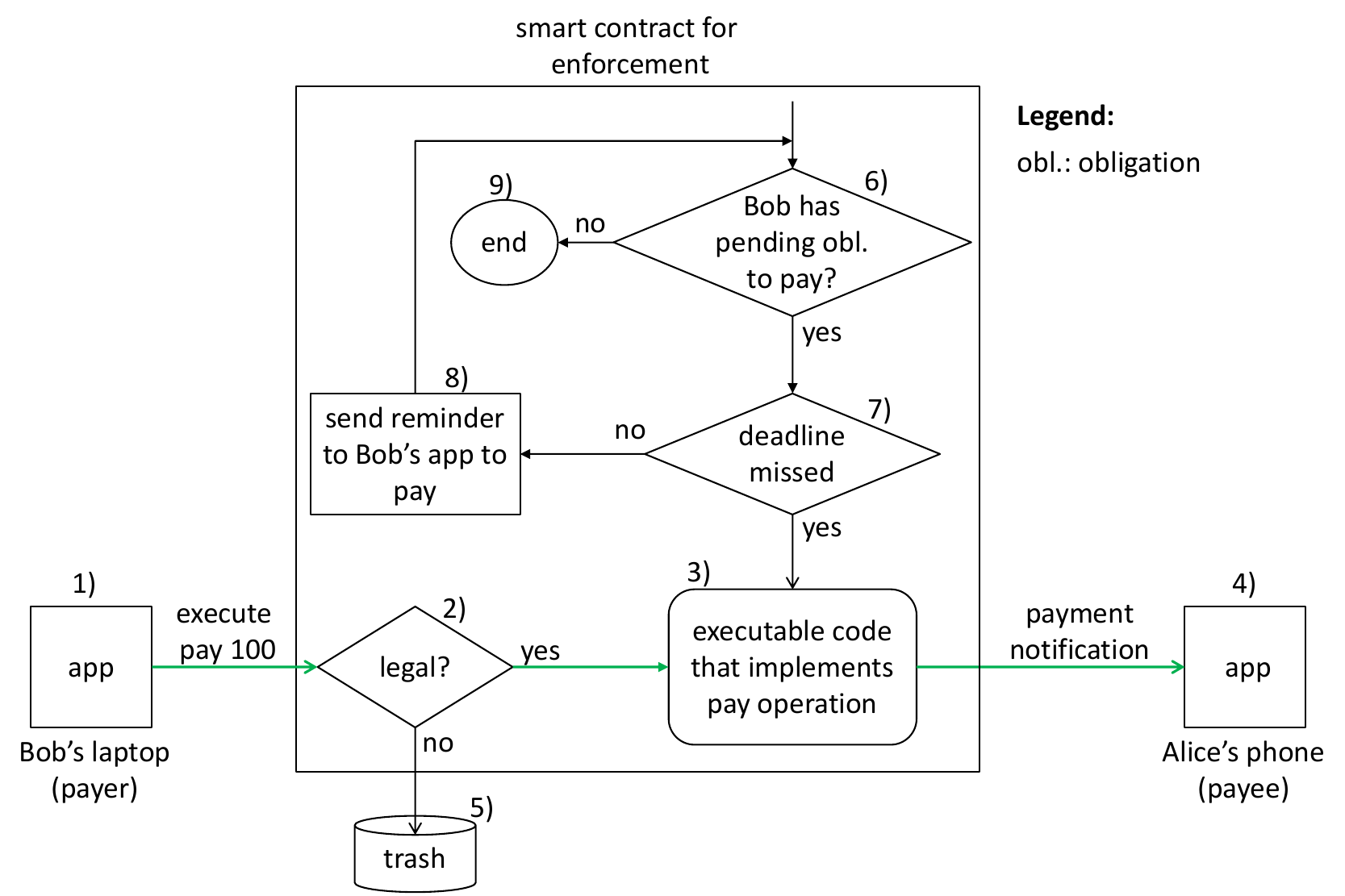}
\caption{Enforcement of the execution of a pay operation}
\label{fig:exepayenforce}
\end{figure}

Contract enforcement is a technique where a smart 
contract is deployed to prevent contract breaches.
As shown in Fig.~\ref{fig:exepayenforce}, to be preventive, 
enforcement operates intrusively (rather than non--intrusively 
like in contract monitoring) in the sense that it interferes with the 
execution of each operation.
 
In the example of the figure, a smart contract is deployed for 
enforcing the execution of the ''pay 100'' operation shown in
Fig.~\ref{fig:exepayoper} and Fig.~\ref{fig:exepaymonitor}.
The smart contract is responsible for ensuring that the ''pay 100'' 
operation is executed as agreed upon, for example, within the 
deadline. Though not shown explicitly in the figure, a Finite 
State Machine (FSM) operates inside the contract. The FSM keeps 
tracks of the current state of the contract, for instance, it 
keeps records about what obligations have been fulfilled. Its 
records can help the contract to determine what operations are 
currently pending and what are legal or illegal, that
is, contract compliant or non--contract compliant.

\begin{enumerate}
\item The ideal execution path is shown by the green line: boxes,
 1--4.
\item Bob's application tries to place the execution of the ''pay 100''
operation against the executable code that implements the pay
operation (box 3).

\item The operation is intercepted (box 2) by the digital contract and 
 analysed for contract compliance.  If it is, the contract 
 forwards the operation to the code that implements the pay 
 operation, otherwise (if the operation is illegal) the
 contract trashes the operation (box 5) so that its execution is
 denied.

 \item If the ''pay 100'' operation reaches the executable
  code, it is executed and Alice is notified.
  Alice's application does not necessarily receive the actual 
  money, it might receive only a payment notification as
  shown in the figure by box 4, for example, a bank receipt.

  \item The enforcement contract is responsible for 
    assuring that Bob complies with his obligation to pay. Accordingly,
    it includes an enforcing mechanism (boxes 6-9) that is programmed
    to send reminders to Bob and to collect the payment if Bob
    fails to honour his obligation.
    
  \item Box 6 checks if Bob has a pending obligation to pay. If the
    answer is ''yes'' the smart contract verifies  (box 7) if the 
    deadline has been missed.
    
 \item If the deadline has not been missed yet, the smart contract
    sends reminders (box 8) to Bob's app.
    
 \item If the deadline has been missed by Bob, the smart contract
    triggers the execution of the executable code that implements
    the pay operation (box 3) to collect Bob's payment  automatically. 
    A pay notification is sent to Alice's app (box 4).

  \item Box 2 allows the execution only of legal operations. For instance, it 
    will not allow Bob's application to execute a ''pay 100'' operation
    outside the pay window, that is, before or after the agreed-upon pay 
    days. Neither will it allow executing ''pay 100''  after the payment 
    has been provided by Bob or enforced by the smart contract.
\end{enumerate}

The power to enforce depends on the particularities 
to execute the operations; for instance, a contract is able to 
enforce Bob's ''pay 100'' operation if it is provided with money in 
advance, such as in escrow, or linked to Alice's accounts via 
some API; otherwise, the contract can only send a warning 
message to Bob's application to remind him of his pending
obligation; next it is up to Bob's application to honour 
or violate the contract.

 \section{Hybrid contracts can provide flexibility}
 \label{hybridcont}
A \textbf{hybrid contract} is a smart contract that is executed and 
enforced automatically by computer executable code in collaboration 
with humans~\cite{LawCommission2020},\cite{LawCommission2021}. Some 
authors use the term \textbf{Ricardian 
contract}~\cite{IanGrigg2015} instead of hybrid and remark that
these contracts consist of two parts that are cryptographically
bound: executable code and tagged text that is human-readable.
Human intervention is required because the assumption is that hybrid 
contracts are incomplete. Thus, at some point, their execution reaches 
a gap left accidentally or intentionally by its designers. 
Incomplete contracts are frequently used in business 
because they offer several advantages, including simplicity and 
flexibility~\cite{Usha2018,Gillian1994,Ian1989}.

We assume that the hybrid contract is designed (drawn up, specified) by 
multidisciplinary professionals with a deep understanding of 
issues that lie at the intersection of fundamental computer science, 
policymaking, jurisprudence and human rights. Also, we assume that 
programmers are responsible for translating the designed contract 
into computer executable code.  We assume that the programmers are 
multidisciplinary professionals with a similar background as the 
designers. We return to this question, Section~\ref{concluding}.

\begin{figure}
\centering
\includegraphics[width=0.55\textwidth]{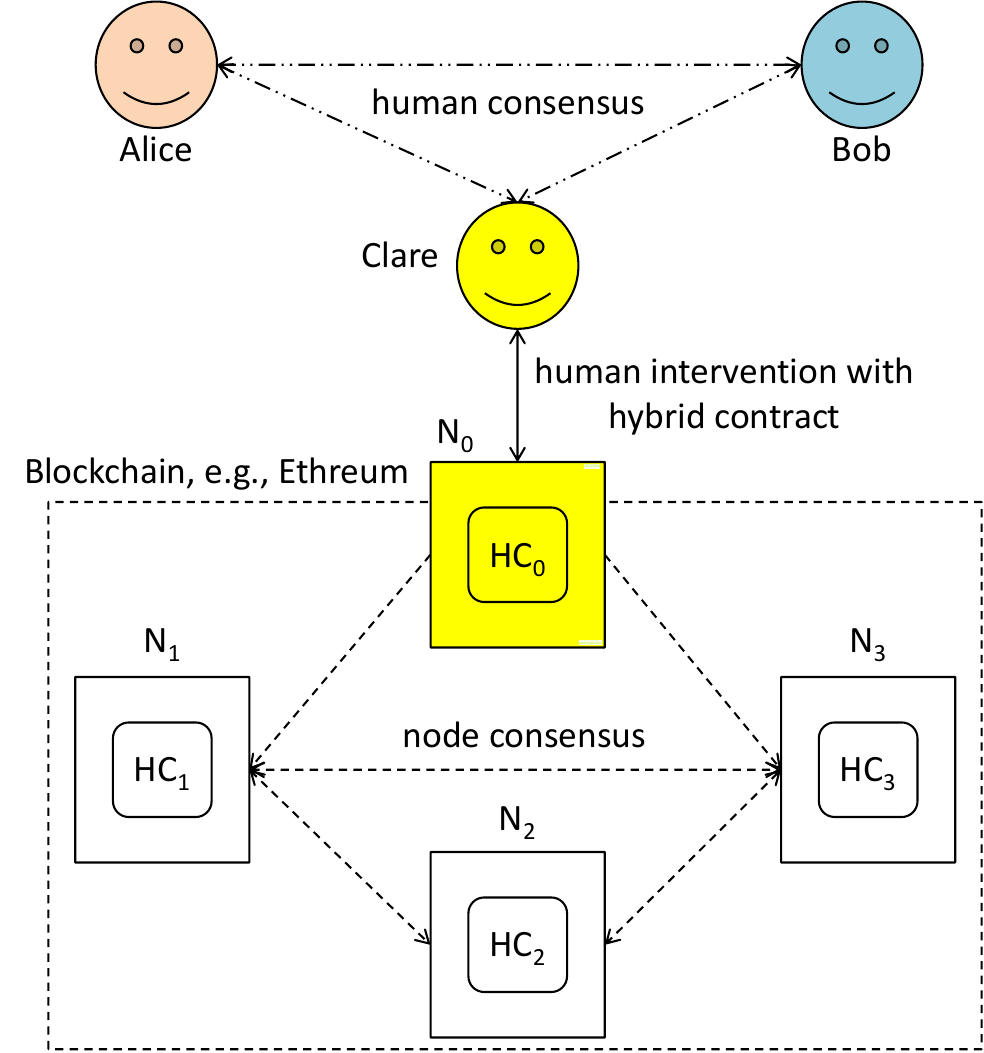}
\caption{A hybrid contract driven  by majority (miners') consensus and human consensus}
\label{fig:hybridcontconsensus}
\end{figure}

 We will use Fig.~\ref{fig:hybridcontconsensus} to explain 
 how a hybrid contract can be used  in algorithmic governance to 
 provide flexibility.
 The figure makes no assumptions about the particularities of law 
 that the hybrid contract is meant to enforce. It can be any legal
 obligation like the enforcement of tax payment, or a right to claim
  insurance or pension, or the enforcement of a prison sentence.
 The bottom part of the figure is the component of hybrid contract that
 executes automatically. It is deployed on a conventional blockchain 
 platform (for example, Ethereum) composed of $N$ nodes that hosts 
 the executable code of the hybrid contract. Only four miner nodes are shown; 
 each of them runs an instance (${HC}_i$) of the executable code. These
 instances are responsible for participating in a consensus protocol
 (say proof of work, proof of stake, etc.) to agree on the execution of 
 an action (called transaction by the blockchain community) that leads to 
 the next state of the hybrid contract.  The figure assumes that the 
 designers intentionally or accidentally left the specification of the hybrid contract 
 incomplete,  thus, at some points it stops and requests human interventions and
 the top part of the figure is activated.  At each stop the decisions to progress 
 the hybrid contract is taken either by a single human (Clare 
 in the figure) unilaterally or, in borderlines situations, by a committee 
 composed of $M$ humans (only three are shown) after running a protocol 
 to reach a consensus. The figure makes no assumptions about the 
 consensus protocol run by Alice, Bob and Clare.

In practice, the automatic (bottom) part of the figure will encode laws 
that are suitable for the majorities and therefore they are likely to be 
enforced (see Fig.~\ref{fig:exepayenforce}). Complementary, the 
non--automatic (top) part of the figure will account for the 
minorities and their odd (exceptional) 
cases that are too risky to solve without human judgment.

For example, imagine that the hybrid contract is responsible for 
enforcing the execution of a prison sentence to be served. An enforcing 
smart contract programmed to operate without a human intervention will act like in
Fig.~\ref{fig:exepayenforce}, that is, impose its algorithm--based
verdict independently. The hybrid contract shown in Fig.~\ref{fig:hybridcontconsensus} 
is more flexible because it will stop to request human intervention to determine 
if the sentence is fair or unfair before enforcing it.

Smart contracts deployed to monitor (see Fig.~\ref{fig:exepaymonitor}) are
inherently flexible because they only observe and collect records about
the activities executed by individuals. The records collected can be examined 
either programmatically by computers or manually by humans. Therefore, the
smart contract is not responsible for making critical decisions. For example,
it is the responsibility of whoever examines the records (not of the smart
contract) to classify a homicide as murder or manslaughter.

\section{Future work and preventive law}
\label{futurework}

Technology provides the opportunity to transform not just 
the judicial system but the legal system in general. We consider 
that the most significant aspect of algorithmic governance does 
not lie in the automation of existing laws to apply them faster, 
at lower costs, or increase efficiency. This is undoubtedly helpful. 
However, in our opinion, the most valuable benefit of algorithmic 
governance is innovation: algorithmic governance opens the 
opportunity to introduce radical changes to the long-standing 
systems that have been used to govern societies for centuries. 
It is widely acknowledged that the democracy that we know suffers 
from numerous shortcomings. The availability of technology 
presents us with the opportunity to include changes that without 
technology were unattainable.

Technology can help algorithmic governance to innovate in 
several fields such as e-democracy, participatory budgets, 
online voting and in the implementation of automatic preventive 
laws. The latter is particularly challenging and a topic in 
our research agenda to progress the discussion presented in 
this paper.  The second author has been studying preventive 
laws as a central topic of her in-progress PhD research. 
Preliminary results are available in an unpublished 
manuscript~\cite{SandraPhD2023}. We will discuss the main 
ideas in this section.

We define automatic preventive laws as an algorithmic system 
that aims at the prevention of the execution
of criminal acts, as opposed to criminal punishment. The 
latter is the prevalent practice in current judicial systems, 
where the perpetrator is punished when the act is already 
committed. We understand that this topic is likely to 
generate controversial discussion because at first glance 
it seems that it attempts against freedom, in fact, careless 
implementation and abuse can result in surveillance and 
repression.
It is important to note that the creation of automatic 
preventive laws is not intended to be repressive.
The challenge is implementing preventive law under the observance 
of human rights as we currently know them and as they might evolve. 

In this context, we use the term law to refer to legal 
norms in general (rules, decrees, wills, and contracts).
These laws would be challenging to circumvent as they are 
applied ex-ante. Additionally, they do not require the 
participation of third parties, such as police or judges 
(who do not always ensure the fairness of
proceedings and the proper administration of justice). 
Consequently, a judicial process would not be necessary. 
Digital tools such as sensors and artificial intelligence 
can help in the development of these systems.

Existing government laws are punitive. They resort to punishing 
criminals and restoring (if possible) the harm
inflicted upon victims. For instance, a law that penalizes 
the actions of a thief who commits a robbery. Digital technology 
can enable the implementation of preventive laws to deter 
individuals from engaging in criminal activities. For example, 
the tax system could automatically collect taxes, eliminating 
both accidental and deliberate tax payers' evasions. A more 
complex and contentious example would be to arrest a potential 
murderer before harming the intended victim.

Preventive law can be used to prevent petty crime (e.g., traffic 
offences and underage drinking). However, their main benefit 
would be preventing catastrophic and often irreversible 
criminal actions, such as murders, government frauds and 
large bank frauds. Reasons for implementing preventive laws 
may lie in the potential to prevent further harm or even 
the loss of innocent lives, which would constitute a 
greater injustice. 

We suggest that automatic preventive laws be gradually 
introduced in various sectors of society and at different 
regulatory levels. However, societies should not adopt automatic 
preventive laws unless they consider human participation 
in extreme circumstances through flexible hybrid smart 
contracts that stops and requests human intervention when necessary.

In our forthcoming analysis, we will explore the application 
of automatic preventative laws to help monitor potential 
monopolistic behaviour online. In a preventive law system, 
artificial intelligence can be used to detect and signal 
if a marketplace manipulates search results to favour 
specific sellers and products. If this happens, corresponding 
legal measures can be enforced automatically before the 
occurrence of illegal acts. This approach ensures fair 
competition and consumer protection. The general idea is 
that automatic preventive laws can act as a powerful 
deterrent against the creation of monopolies.

\section{Conclusions}
\label{concluding}
Algorithmic governance is all about law automation, which is 
an alternative that has the potential to ameliorate several 
problems that afflict current legal systems, such as its 
unacceptable slow pace and lack of impartiality introduced 
by judges that succumb to corruption. However, if technology 
is adopted, care should be taken not to create a 
computer--driven system that is unnecessarily rigid and 
de--humanized. Automated law should not be embraced unless 
it accounts for humans involvement in border--line situations 
where human intelligence is likely to produce fairer decisions. 
 
Some authors refer to algorithmic governance as Lex 
Algomata~\cite{LeyAlgomata} to emphasize the risk that 
algorithmic governance can bring. Algomata is a term that 
possesses profound meaning and is formed by combining 
two words: the word \emph{algo} which is the prefix of 
algorithm, and the word \emph{mata} which in English 
means to kill and is also the suffix of the word 
automata\footnote{An automaton is the graphical representation 
of an algorithm.}. Therefore, \emph{algomata} carries a 
negative connotation and emphasizes that careless use 
of automation in the field of Law can have undesirable 
and irreversible consequences.
 
Let us not forget ancient traditions and the human factor in 
discussions of law governance. It is essential to bear in mind 
that the transformation of legal practice by the use of automation 
of law will not be necessarily welcome by the lawyer community. 
One of the problems is that Law automation  can lead to job 
losses for many legal professionals who do not have 
sufficient knowledge of technology. However, it may be 
a good opportunity for younger or newly graduated 
attorneys who adapt faster to technological changes.

A further important point to consider is the opportunities 
and challenges that the automation of law will bring to courts, 
law firms and lawyers and people who need their services. 
In relation to procedural law, on the one hand, the vision is that
automated law will help judges to resolve cases that need minimal human 
intervention in less time.  But on the other hand, it will force judges
to become familiar with programming languages to 
understand code and, in general, with computer technology. A 
more general and fundamental question here 
is which professionals will be responsible for the 
implementation of the programs (say, the smart contracts) 
which are needed to automate the law, to certified that 
they are correct and to interpret their results when 
human intervention is needed. We are asking for multidisciplinary 
professionals with Law and Computer Science backgrounds, that 
is, lawyers with knowledge to read programming code and 
software engineers with the knowledge to read civil 
codes~\cite{JamesGrimmelmann2022}. Currently, such professionals 
are missing, and it is not clear who and where they can be 
trained. A possible solution is to create a branch of
traditional software engineering to cover the existing 
gap~\cite{SandraCarlosDina2022}. In fact,
some authors have already suggested the creation 
of Blockchain--Oriented 
Software Engineering (BOSE)~\cite{Mahdi2021},\cite{Giuseppe2018},
 \cite{Porru2017}. We agree with their views but suggest 
that BOSE covers legal aspects too
 and thoroughly. These professionals will help to design 
and program the smart contracts and to react to requests 
for human intervention placed by hybrid contracts.

 Another important issue is that total automation can 
 generate rigidity in a system and inflexibility in 
 decision--making. For this reason and as argued 
 in this paper, we believe that sometimes unilateral 
 decisions of a computer system or a single individual do 
 not offer a fair solution. In these situations, consensus 
 that emerges from the agreement between several people 
 could provide greater certainty about decisions.

 Let us not forget that so far, technologists have not 
 been able to produce technology that is 100\% reliable. 
 Their current technology is embarrassingly brittle to 
 rely on it for serious matters such as dictation of a 
 prison sentence or, in some countries, executions.

The recent outage of Facebook, WhatsApp and Instagram 
on 4 Oct 2022 that lasted for about six hours can help 
to illustrate the argument.
Apparently, the outage left billions of users without the
services was caused by an internal configuration 
issue\footnote{\url{https://www.bbc.co.uk/news/technology-58793174}}
and suggests that failures are unavoidable. As a second example,
we can mention the failures AI technologies that have been already
used to assist in law automation. There are examples that
have shown that AI algorithms are not infallible. For example, facial 
recognition used in the criminal sphere has led to biases (i.e., racial) 
decisions. This example shows that careless use of 
unsound technology can result in systematic discrimination of 
minority groups~\cite{Samer2017},\cite{CharlesChoi2021},
\cite{Natalianorodi2021}.

Perfecting this technology will take time. Therefore, we consider 
that while these technological problems are solved, it is not convenient 
to fully automate or make decisions based entirely on the orders issued 
by a machine. Human intervention is essential to protect the life, 
equality, dignity and fundamental rights of people.

\begin{Backmatter}

\paragraph{Acknowledgments}
Carlos Molina--Jimenez has been supported by UKRI grant G115169.
For the purpose of open access, the author Carlos 
Molina-Jimenez has applied a Creative Commons Attribution 
(CC BY) licence to any Author Accepted Manuscript version 
arising from this submission.

\paragraph{Funding Statement}
The first author was supported by the Department of Computer 
Science and Technology of the University of Cambridge 
(www.cst.cam.ac.uk) and TODAQ (todaq.net).
 
\newpage  
\bibliographystyle{apalike}  
\bibliography{bibliography}
 
\end{Backmatter}
\end{document}